\documentclass[11pt]{article} 
\usepackage[paper=a4paper,margin=1.0in]{geometry}

\usepackage{amssymb,amstext,amsmath,amsthm,mathtools}
\usepackage[dvips]{graphicx}
\usepackage{latexsym}
\usepackage{psfrag}
\usepackage{amsfonts}
\usepackage{bm} 
\usepackage{color}
\usepackage{cite}
\usepackage[colorlinks=true,linkcolor=blue,citecolor=red,linktocpage=true,dvipdfmx]{hyperref}
\usepackage[utf8]{inputenc} 
\usepackage{enumitem}
\numberwithin{equation}{section}

\def\nn{\nonumber}
\def\lb{\label}
\def\ci{\cite}

\newcommand{\p}{\partial}
\newcommand{\N}{\nabla}
\def\cL{ {\cal L}}

\def\a{\alpha}
\def\b{\beta}
\def\g{\gamma}
\def\d{\delta}
\def\D{\Delta}

\def\e{\epsilon}

\def\s{\sigma}

\def\bra{\langle}
\def\ket{\rangle}
\def\l{\left}
\def\r{\right}
\def\f{\frac}
\def\MO{\mathcal {O}}
\def\MR{\mathcal {R}}
\def\MH{\mathcal {H}}

\begin{document}
\title{Self-Gravity and Bekenstein-Hawking Entropy
} 
\author{Yuki Yokokura \footnote{yuki.yokokura@riken.jp}
\\
{\small \emph{iTHEMS Program, RIKEN, Wako, Saitama 351-0198, Japan}}}
\date{}

\maketitle
\begin{abstract}
We study the effect of self-gravity on entropy by directly solving the 4D semi-classical Einstein equation. In particular, we focus on whether the Bekenstein-Hawking formula holds when self-gravity is extremely strong. As an example, we consider a simple spherically symmetric static configuration consisting of many quanta and construct a self-consistent non-perturbative solution for $\hbar$ in which the entropy exactly follows the area law for many local degrees of freedom of any kind. This can be a candidate for black holes in quantum theory. It represents a compact dense configuration with near-Planckian curvatures, and the interior typically behaves like a local thermal state due to particle creation. Here, the information content is stored in the interior bulk, and the self-gravity plays an essential role in changing the entropy from the volume law to the area law. We finally discuss implications to black holes in quantum gravity and a speculative view of entropy as a gravitational charge. 
\end{abstract}

\section{Introduction.}\lb{s:intro}
Information has weight. 
Physically, information is represented by an excited quantum state $|\psi\ket$, 
and the excitation energy $\bra \psi|T_{\mu\nu}|\psi \ket$ generates self-gravity as a curved spacetime $g_{\mu\nu}$. 
This is described by the 4D semi-classical Einstein equation: 
\begin{equation}\lb{Einstein}
G_{\mu\nu}=8\pi G \bra \psi|T_{\mu\nu}|\psi \ket.
\end{equation}

As a result, the amount of information, thermodynamic entropy $S$, depends on self-gravity. 
This is a manifestation of the unshielded long-range nature of gravitational force, 
while other gauge fields have effective short-range interactions due to Debye shielding \ci{Gross,Pad_thermo,Landau_SM}. 
Suppose, say, a spherical static object with size $R$ and finite energy-momentum distribution, whose exterior is almost vacuum.
This object can be considered as a collection of excited quanta in $(g_{\mu\nu},|\psi\ket)$ satisfying \eqref{Einstein}. 
The entropy $S$ of the system can be expressed phenomenologically as 
\begin{equation}\lb{S}
S=\int^R_{0} dr \sqrt{g_{rr}(r)}s(r),
\end{equation}
where  $s(r)$ is the entropy density per proper radial length. 
In a short range compared to the radius of curvature, 
thermodynamics should hold locally because of the equivalence principle, 
and the entropy density per proper volume, $\f{s(r)}{4\pi r^2}$, should be 
an increasing function of the local temperature $T_{loc}(r)$ \ci{Landau_SM}. 
In a longer range, however, gravity makes the region non-uniform, 
and the extensivity of thermodynamic quantities breaks down, 
leading often to negative heat capacity \ci{Landau_SM,Lynden}. 
Indeed, the proper length factor $\sqrt{g_{rr}(r)}$ appears in \eqref{S}, 
and \textit{if} Tolman's law \ci{Tolman,Landau_SM} holds properly, 
the local temperature $T_{loc}(r)$ can be different at points depending on $\sqrt{-g_{tt}(r)}$. 
Thus, the entropy of self-gravitating systems is determined by the metric $g_{\mu\nu}$ satisfying \eqref{Einstein}. 
An example is self-gravitating thermal radiation in a spherical box of size $R$; 
while the entropy density follows the Stefan-Boltzmann law locally, the total entropy $S$ 
is proportional to $R^{3/2}$ not $R^3$ as a result of \eqref{Einstein} \ci{Sorkin}. 

An interesting fundamental question arises here: \textit{Can we increase the strength of self-gravity without forming a horizon and construct a configuration whose entropy follows the Bekenstein-Hawking formula \ci{Bekenstein,Hawking}?} Note here that the value of the formula is determined by conserved charges such as ADM energy and does not depend on the presence of a horizon. 
The question should be important for studying 
quantum black holes and the relation between self-gravity and the entropy bounds conjectured \ci{B_bound,Bousso1}, which ultimately should connect to searching for degrees of freedom of quantum gravity \ci{tHooft, Susskind, Bousso2}. 
In this paper, we solve directly \eqref{Einstein} and consider the effect of self-gravity on entropy to construct a self-consistent configuration that satisfies the Bekenstein-Hawking formula.\footnote{
Similar problems are addressed, say, in Refs.\ci{Oppenheim,Pesci}, where the existence of matter satisfying Tolman's law globally and a certain energy-momentum tensor is assumed a priori, but it is not discussed whether it really satisfy the Einstein equation self-consistently. On the other hand, we actually construct a self-consistent solution that satisfies the area law and  \eqref{Einstein}, including an evaluation of the renormalized energy-momentum tensor $\bra\psi|T_{\mu\nu}|\psi\ket$.} 

As a simple trial,
we consider a spherical static configuration with ADM mass $M=\f{a}{2G}$, 
consisting of quanta uniformly distributed with respect to the proper radius, 
except for a small region around $r=0$ where the semi-classical approximation may break down ($a\gg l_p\equiv \sqrt{\hbar G}$). 
That is, they are uniformly distributed in $l_p\ll r \leq R$,
where the size $R$ is larger than but close to $a$. 
These quanta are responsible for the information $|\psi\ket$ and thus the entropy. 

To express the strength of gravity in such a static configuration, we can use, from the equivalence principle, 
the proper acceleration $\a_n(r)$ required to stay at $r$ inside.  
We here note two observations. 
First, in the Schwarzschild metric with $M=\f{a}{2G}$, 
$\a_n(r)$ increases as $r\to a$ 
and becomes $\MO(1/l_p)$ at $r\sim a+l_p^2/a$ \ci{Poisson},
where $\D r\sim l_p^2/a$ corresponds 
to the typical fluctuation of the mass, $\D M\sim T_H=\f{\hbar}{4\pi a}$. 
Second, in quantum gravity, the limitation of spacetime resolution is considered to be the Planck length $l_p$ \ci{Pad,string,Loop,Yoneya}, 
which is related to the existence of the maximum acceleration $\sim 1/l_p$ \ci{Cai1,Brandt,RV}. 
Therefore, the characteristic scale at $r$, given by $\a_n(r)^{-1}$, 
must be sufficiently longer than $l_p$ for the semi-classical description to hold. 
Motivated by these, we thus characterize the uniform configuration 
with the semi-classically maximum gravity by 
$\a_n(r)\approx{\rm const.}=\MO(1/{\cal C} l_p)$ 
for $l_p\ll r \leq R$, where ${\cal C}$ is a large number of $\MO(1)$ to be determined.\footnote{
$\MO(1)$ means $\MO(r^0)$ or $\MO(a^0)$ for $r,a\gg l_p$.}

\begin{figure}[h]
\begin{center}
\includegraphics*[scale=0.11]{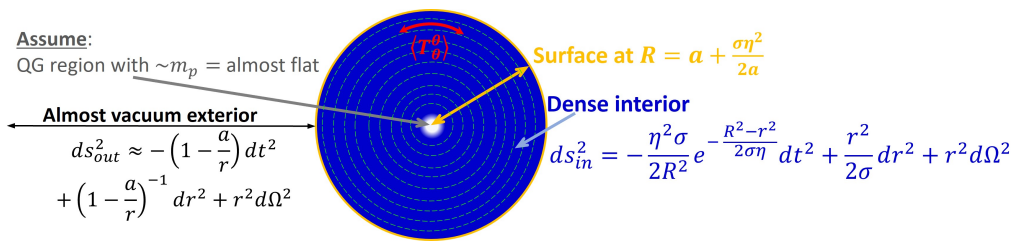}
\caption{\footnotesize The dense configuration with mass $M=\frac{a}{2G}$ and the Bekenstein-Hawking entropy. Continuous concentric excited quanta responsible for the entropy are uniformly distributed in the proper radial length.}
\label{f:Whole}
\end{center}
\end{figure}

Using these two conditions, 
we construct the interior metric of the configuration as a self-consistent solution of \eqref{Einstein}.  
The key is to implement the notion ``weight of information": 
we use a mechanical argument motivated by Bekenstein's thought experiment \ci{Bekenstein}, 
estimate the excitation energy of the uniformly distributed quanta responsible for the entropy, 
and express the entropy density $s(r)$ in terms of the metric, 
to construct the metric consistent with the area law and the acceleration condition. 
Then, the interior metric for $l_p\ll r\leq R$ becomes
\begin{equation}\label{interior_I}
   ds^2=-\f{\s\eta^2}{2R^2}e^{-\f{R^2-r^2}{2\s\eta}}dt^2+\f{r^2}{2\s}dr^2+r^2d\Omega^2,
\end{equation}
where two parameters $\sigma=\MO({\cal C}^2 l_p^2)$ and $\eta=\MO(1)$ are determined by solving \eqref{Einstein} self-consistently. See Fig.\ref{f:Whole}. 
This has near-Planckian curvatures, which induce vacuum fluctuations of various modes to produce a large tangential pressure $\bra \psi|T^\theta{}_\theta|\psi\ket$, and the interior is largely anisotropic, which allows the configuration to exceed Buchdahl's bound \ci{Buchdahl}. 
The exterior metric for $r\geq R$ is approximately given by the Schwarzschild metric with mass $\f{a}{2G}$. Both are connected smoothly at $r=R$, where 
\begin{equation}\label{R_I}
    R=a+\frac{\s\eta^2}{2a}
\end{equation} 
is fixed consistently with Israel's junction condition \ci{Poisson}. This is close to the Schwarzschild radius $a$, and the object looks like a classical black hole from the outside; indeed, the information inside is captured for a long time due to the exponentially large redshift in \eqref{interior_I} although it has no horizon. 
The center small region ($0\leq r \lesssim {\cal C} l_p$), beyond the semi-classical approximation, has only a small energy $\sim m_p(\equiv\sqrt{\hbar/G})$ and can be expected to be a small excitation of quantum gravity degrees of freedom. 
Thus, the interior metric \eqref{interior_I} represents a compact dense configuration as in Fig.\ref{f:Whole} consisting of the excited quanta and the vacuum fluctuations, with the surface at $r=R$ and the large finite curvatures, instead of a horizon or singularity. 
This is a non-perturbative solution in $\hbar$ to the 4D semi-classical Einstein equation \eqref{Einstein} with many local degrees of freedom. 

In the self-consistent state $|\psi\ket$, the excited quanta behave typically like a local thermal state at a near-Planckian local temperature $T_{loc}(r)$ due to particle creation inside,
leading to the equilibrium with a heat bath at the Hawking temperature $T_H=\f{\hbar}{4\pi a}$. 
Evaluating the entropy density $s(r)$ through thermodynamic relations and summing it up over the interior volume, 
the total entropy \eqref{S} agrees with the Bekenstein-Hawking formula including the factor $1/4$ for any type of local degrees of freedom. 
The strong self-gravity changes the entropy from the volume law to the area law, while the information itself is stored in the interior bulk. Thus, this non-perturbative configuration without horizon can be a candidate for quantum black holes (at least, at a semi-classical level) in the sense that the entropy follows the Bekenstein-Hawking formula. This is also an example of having no horizon and saturating the entropy bounds \ci{B_bound,Bousso1}.

Here, we comment on the motivation for our two assumptions: uniformity and maximum acceleration. 
First, it should be natural to start with simplest possible models to investigate new physics; historically, for example, uniform models have been studied when examining the interiors of atoms and stars. 
Second, these two assumptions are, logically, only sufficient conditions for the Bekenstein-Hawking formula, 
and it is not clear yet whether they are necessary conditions. 
On the other hand, there is an interesting fact: 
if thermal radiation is collected reversibly by self-gravity in a heat bath at Hawking temperature, 
a configuration satisfying the two assumptions can be obtained as a result of time evolution in a simple model according to \eqref{Einstein} \cite{KMY,KY1}.
(See Appendix \ref{A:M_model} for a review.) Thus, from a thermodynamic standpoint, 
the two assumptions should be physical and interesting.

The argument will be developed in a \textit{self-consistent manner}. 
First, we find a \textit{candidate} metric such that the entropy is proportional to the area law (Sec.\ref{s:metric}): we use $s(r)=const.$ (from the uniformity), $\alpha_n(r)=\MO(1/{\cal C}l_p)$ (from the maximum acceleration), and $S\propto a^2/l_p^2$ (from the consistency to the area law), to obtain the metric \eqref{interior_I} for a given $(\s,\eta)$. 
Then, we study the behavior of quantum fields in this background metric. 
We apply the Unruh effect locally to the metric to obtain the local temperature $T_{loc}(r)$ (Sec.\ref{s:thermal}). 
We then use thermodynamic laws locally to derive the area law exactly (Sec.\ref{s:entropy}). 
We determine the position $R$ of the surface as \eqref{R_I} from a thermodynamic equilibrium condition in a heat bath (Sec.\ref{s:surface}). 
Now, using conformal matter fields as an example, 
we evaluate the renormalized energy-momentum tensor $\bra \psi|T_{\mu\nu}|\psi \ket$, equate it to $G_{\mu\nu}$ and determine the self-consistent values of $(\s,\eta)$, to show that the metric \eqref{interior_I}  satisfies \eqref{Einstein} self-consistently, where the effect of the vacuum fluctuations is crucial (Sec.\ref{s:self}). 
Finally, we consider the time evolution of the quantum fields 
in the formation and evaporation process of this configuration  
and derive that particles are dynamically created inside at the same temperature $T_{loc}(r)$ (Sec.\ref{s:T}). 
Thus, we obtain the self-consistent configuration satisfying the Bekenstein-Hawking formula and \eqref{Einstein}.
We conclude with a summary, relations to the previous works \cite{KMY,KY1,KY2,Ho,KY3,KY4,KY5}, comparisons with other horizon/singularity avoiding scenarios, and future directions: information recovery from the interior structure, entropy including self-gravity as a gravitational charge, and implications to black holes in quantum gravity 
(Sec.\ref{s:con}). 

\section{Interior metric from information}
\begin{figure}[h]
\begin{center}
\includegraphics*[scale=0.12]{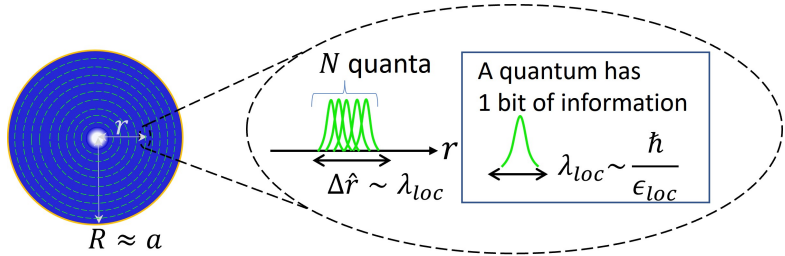}
\caption{\footnotesize A microscopic picture of the configuration 
consisting of uniformly distributed quanta with 1-bit of information and local energy $\e_{loc}$.}
\label{f:picture}
\end{center}
\end{figure}

\lb{s:metric}
We construct a candidate interior metric for $l_p\ll r\leq R$, which can be set as
\begin{equation}\lb{metric}
ds^2=-e^{A(r)}dt^2 + B(r)dr^2 + r^2d\Omega^2,
\end{equation}
while the exterior one for $r\geq R$ is discussed in Sec.\ref{s:surface}. 
Suppose that there are $N$ quanta with local energy $\e_{loc}$ 
and 1 bit of information in a width $\D \hat r$ around $r$ (see Fig.\ref{f:picture}).
Here, $N$ is a number to be determined, the meaning of 1-bit of information is described below, 
and $(\hat t,\hat r)$ is the local coordinate with $\D \hat t=\sqrt{-g_{tt}(r)}\D t$ and $\D \hat r = \sqrt{g_{rr}(r)}\D r$.
Then, the total local energy, $N \e_{loc}=4\pi r^2 \D \hat r \bra T^{\hat t\hat t} \ket$, 
leads to 
\begin{equation}\lb{e_den}
-\bra T^t{}_t(r)\ket=\f{N\e_{loc}}{4\pi r^2 \D \hat r},
\end{equation}
where we have used $g^{\hat t\hat t}=-1$ and $T^t{}_t=T^{\hat t}{}_{\hat t}=-T^{\hat t\hat t}$. 
Here, the contribution to the ADM energy of the part within $r$ in a spherically symmetric system (or the Misner-Sharp mass) is given through \eqref{Einstein} by 
\begin{equation}\lb{ADM}
M(r)=4\pi \int^r_0 dr' r'^2 \bra-T^t{}_t(r')\ket,
\end{equation}
where $\lim_{r\to \infty}M(r)=M$, and $(r^2\bra T^t{}_t(r)\ket)|_{r\to 0}$ is finite \ci{Landau_C,Poisson}. 
Combining \eqref{e_den} and \eqref{ADM} yields 
the formula for the ADM energy $\D M_{1bit}$ of a 1-bit quantum located at $r$: 
\begin{equation}\lb{dM1}
\D M_{1bit}= 4\pi r^2 \D r \f{\bra - T^t{}_t(r)\ket}{N}=\f{\e_{loc}}{\sqrt{g_{rr}(r)}}.
\end{equation}

On the other hand, by using the uniformity and Bekenstein's argument, 
we can obtain 
\begin{equation}\lb{dM2}
\D M_{1bit} \sim \f{\hbar}{r}. 
\end{equation}
The reason is as follows. 
First, the total size $R$ will be determined later as $R=a+\MO(l_p^2/a)$ (see \eqref{R}), 
which is of the same order as the radius at which Bekenstein's thought experiment is performed \ci{Bekenstein}. 
Second, the physical property at any radius $r$ is equivalent due to the uniformity in the radial direction, 
and the interior of a given $r$ is not affected by its exterior due to the spherical symmetry. 
Therefore, the size of the configuration with mass $M(r)$ is $r=a(r)+\MO(l_p^2/a(r))$,
where $a(r)\equiv2GM(r)$.
Then, applying Bekenstein's idea to massless fields, a quantum with $\D M \sim \f{\hbar}{r}$ has wavelength $\sim r$ and carries 1 bit of information about whether it would enter or not when forming the configuration of mass $M(r)$ and size $\approx r$: the wavelength and the size of the configuration are comparable, and the probability of entering is about $1/2$, which means that the entropy is 1-bit and thus the quantum has the 1-bit of information about whether it exists there or not.  
Hence, from the continuity of the uniform distribution, a quantum with the 1-bit of information at $r$ has energy \eqref{dM2}. 

Thus, by equating \eqref{dM1} and \eqref{dM2}, 
we can evaluate the proper wavelength as 
$\lambda_{loc}\sim \f{\hbar}{\e_{loc}} \sim \f{r}{\sqrt{g_{rr}(r)}}$.
Now, there are $N$ quanta with 1 bit of information and this wavelength 
in the width $\D \hat r \sim \lambda_{loc}$. 
Therefore, the entropy per unit proper length $s(r)$ can be estimated as
\begin{equation}\lb{s_def}
s(r)\sim \f{N}{\D \hat r}\sim \f{N\sqrt{g_{rr}(r)}}{r}.
\end{equation}
Because the uniform spherical distribution means that $s(r)$ is constant, 
\eqref{s_def} requires us to set $g_{rr}(r)=\f{r^2}{2\s}$, where $\s$ is a constant. 
Then, the total entropy \eqref{S} can be estimated as
\begin{equation}\lb{S_eva}
S=\int^R_{\sim l_p} dr \sqrt{g_{rr}(r)}s(r) \sim \int^a_0 dr \f{N g_{rr}(r)}{r}\sim \f{N a^2}{\s}.
\end{equation}
For this to be consistent with $S\sim \f{a^2}{l_p^2}$, 
we must have\footnote{Note that this requirement is just the first step in the self-consistent argument, 
and at this stage we don't claim that the area law is derived.}
\begin{equation}\lb{B}
g_{rr}(r)=B(r)=\f{r^2}{2\s},~~\s\sim N l_p^2.
\end{equation}

As a result, the entropy density \eqref{s_def} becomes 
$s(r)\sim \f{\sqrt{\s}}{l_p^2} \sim \f{\sqrt{N}}{l_p}$.
Thus, $\s$ determines the entropy density, and 
$\MO(\sqrt{N})$ bits of information are packed per the proper Planck length. 
Note that the wavelength and local energy for 1 bit of information 
are estimated from \eqref{B} 
as 
\begin{equation}\lb{wave}
\lambda_{loc}\sim \sqrt{N}l_p,~\e_{loc}\sim \f{m_p}{\sqrt{N}}, 
\end{equation}
respectively, where $m_p\equiv \sqrt{\f{\hbar}{G}}$. 

Next, we fix $A(r)$ by considering the other condition, $\a_n(r)=\MO\l(\f{1}{{\cal C}l_p}\r)$, 
where $\a_n(r)\equiv |g_{\mu\nu}\a^\mu_n\a^\nu_n|^{\f{1}{2}}$, $\a^\mu_n\equiv n^\nu \N_\nu n^\mu$ 
and $n^\mu \p_\mu =(-g_{tt}(r))^{-\f{1}{2}}\p_t$.
This can be expressed in \eqref{metric} as 
\begin{equation}\lb{an}
\a_n(r)=\f{\p_r \log \sqrt{-g_{tt}(r)}}{\sqrt{g_{rr}(r)}}
=\f{1}{\sqrt{2\s \eta^2}},
\end{equation}
by introducing another constant parameter $\eta$ satisfying $N \eta^2=\MO(1)\gg 1$. 
Integrating this through \eqref{B} provides 
\begin{equation}\lb{A}
A(r)=\f{r^2}{2\s \eta}+A_0,
\end{equation}
where $A_0$ is an integration constant to be fixed later. 
Thus, we reach the candidate metric: 
\begin{equation}\lb{interior0}
ds^2=-e^{\f{r^2}{2\s \eta}+A_0}dt^2 + \f{r^2}{2\s}dr^2 + r^2d\Omega^2,
\end{equation}
which is \eqref{interior_I} up to the constant $A_0$. 
The two parameters $(\s,\eta)$ can be determined by solving \eqref{Einstein} self-consistently (see Sec.\ref{s:self}).
Note that the above procedure provides a method of constructing a metric 
from information such as a total entropy $S$ and a distribution $s(r)$. 

Let us now study the properties of the interior metric \eqref{interior0}. 
If \eqref{Einstein} is applied, 
we have 
\begin{align}\lb{Ttt}
-\bra T^t{}_t(r)\ket &=\f{1}{8\pi G r^2}, \\
\lb{Trr}
\bra T^r{}_r(r)\ket &= \f{2-\eta}{\eta} (-\bra T^t{}_t(r)\ket), \\
\lb{T33}
\bra T^\theta{}_\theta(r)\ket &= \f{1}{16\pi G \s \eta^2},
\end{align}
as the leading terms for $r\gg l_p$. 
First, the energy density \eqref{Ttt} reproduces through \eqref{ADM} mass $M=\f{R}{2G}\approx \f{a}{2G}$ 
if the region outside $r=R\approx a$ is almost vacuum.
Second, the tangential pressure \eqref{T33} is almost Planckian 
and makes the interior anisotropic ($\bra T^\theta{}_\theta\ket \gg \bra T^r{}_r\ket$), 
which allows avoiding Buchdahl’s limit \cite{Buchdahl}. 
This pressure stabilizes the configuration against the self-gravity, 
which can be checked in the anisotropic TOV equation, and violates the dominant energy condition. 
In this sense, we can expect that the origin of \eqref{T33} is different from that of \eqref{Trr}. 
We will see in Sec.\ref{s:self} that this is the case.
Third, in a static gravitational field, 
high excitations of local degrees of freedom, like \eqref{wave}, should be locally consistent with thermodynamics, 
and pressures should be positive and not larger than energy density \ci{Landau_SM,Landau_C}. 
From the expectation above, this should be applied only to the radial pressure  \eqref{Trr}, requiring that 
\begin{equation}\lb{eta}
1\leq \eta <2.
\end{equation}
To hold $N \eta^2=\MO(1)\gg 1$, therefore 
$N$ must be a constant of $\MO(1)$ much larger than 1: 
\begin{equation}\lb{large_N}
N \gg1.
\end{equation}
Thus, the metric \eqref{interior0} represents a dense object with size $R$ 
in that the energy density is finite inside and the tangential pressure is large. 

The leading values of the curvatures for $r\gg l_p$ 
are semi-classically maximum:
\begin{align}\lb{RRR}
\MR =-\f{2}{L^2},~~R_{\mu\nu}R^{\mu\nu}&=\f{2}{L^4},~~R_{\mu\nu\a\b}R^{\mu\nu\a\b}=\f{4}{L^4}, \\
\lb{L}
L&\equiv \sqrt{2\s\eta^2} \sim \sqrt{N}l_p \gg l_p,
\end{align}
which are constant and thus consistent with the uniform condition. 
We note three points here. 
One is that the configuration should have no singularity. 
Indeed, if we apply \eqref{ADM} and \eqref{Ttt} to $r\sim\sqrt{N}l_p$, 
such a small semi-classical region has the energy $M(r\sim \sqrt{N}l_p)\sim \sqrt{N} m_p$. 
Therefore, the center region $0\leq r \lesssim l_p$ is expected 
to have at most only a small energy $\sim m_p$
and be described as a quantum-gravity regular state, say, an excitation of string.
In this sense, the center can be considered almost flat,\footnote{
In a simple model, we can see explicitly that the center is kept flat as a result of the semi-classical time evolution of a spherical collapsing matter (except for the last moment of the evaporation) \ci{KMY,KY4} (see also Appendix \ref{A:M_model}).} 
and the metric \eqref{interior0} describes the region $\sqrt{N}l_p \lesssim r \leq R$.
Another one is, 
from the Ricci scalar $\MR$ in \eqref{RRR},\footnote{
More precisely, the Ricci scalar is $\MR=-\f{2}{L^2}+\MO(r^{-2})$.} 
that the metric \eqref{interior0} is a warped product of $AdS_2$ of radius $L$ \eqref{L} and $S^2$ of radius $r$ 
(see \ci{KY4} for an explicit proof). 
The last one is that the curvature radius $L$ \eqref{L} is almost the same length 
as the wavelength $\lambda_{loc}\sim \sqrt{N}l_p$ (from \eqref{wave}), 
and thus the above arguments based on the local frame $(\hat t,\hat r)$ are valid, albeit barely.


\section{Local thermal behavior}\lb{s:thermal}
The interior metric \eqref{interior0} originates from the self-gravity of the quanta responsible for the entropy 
and other modes that may be induced self-consistently. 
Let us here study the behavior of the former ones in this background metric. 

They are in uniformly accelerated motion with acceleration \eqref{an} against the gravity 
such that they are steadily distributed on a $t$-constant hypersurface. 
This distribution is covariant in that the quanta are stationary 
with respect to the timelike Killing vector $\p_t$, a covariant notion. 

We now consider the Unruh effect locally \ci{Unruh, BD}. 
In general, in a local spacetime region of size smaller than the radius $\cL$ of the curvature, 
a quantum with an acceleration $\a \gtrsim 1/\cL$ feels the temperature $T_{U}=\f{\hbar \a}{2\pi}$ \ci{Jacobson,Pad2}, 
although the effect may be negligible compared to the excitation energy. 
In our case, the acceleration \eqref{an} is barely on the same scale as the curvature \eqref{RRR} ($\a_n(r)=1/L\sim 1/\cL$), 
and the quanta in the local region with the coordinate system $(\hat t,\hat r)$ 
feel the temperature $T_{U}(r)=\f{\hbar \a_n(r)}{2\pi}=\f{\hbar}{2\pi L}\sim \f{m_p}{\sqrt{N}}$. 
Notably, since this temperature is on the semi-classically maximum energy scale \ci{Pad},  
the quanta in any initial state $|\psi\ket$ will become like a thermal state after transitioning to this configuration \ci{Caianiello}. 
Therefore, the quanta at each $r$ typically behave like the local thermal state in the radial direction 
at the local temperature 
\begin{equation}\lb{T_loc}
T_{loc}(r)=\f{\hbar}{2\pi L},
\end{equation}
which is consistent with the energy scale for 1 bit, $\e_{loc}\sim \f{m_p}{\sqrt{N}}$ \eqref{wave}. 
In Sec.\ref{s:T}, this will be proved at a dynamical level. 
Note that this holds 
for any type of local degree of freedom because of the universality of the Unruh effect.

We here discuss the consistency with Tolman's law \ci{Tolman,Landau_SM}. 
First, the fact that each small part of the interior 
has the constant local temperature \eqref{T_loc} 
might at first glance appear to violate Tolman's law. 
In general, the law holds only if thermal radiation can propagate between objects at rest with respect to each other 
in a stationary spacetime within a time considered \ci{Tolman}. 
For the dense configuration, 
due to an exponentially large redshift (see \eqref{interior}), 
the small parts separated by a distance $\D r =\MO(a)$ 
cannot exchange radiation until time $\D t= \MO(e^{a^2/l_p^2})$ has passed (see \ci{KY5} for details). 
Therefore, only during $\D t\ll \MO(e^{a^2/l_p^2})$, 
can the configuration exist in radially local (not global) equilibrium, 
consistent with Tolman's law.\footnote{This local equilibrium makes the interior different 
from that of \ci{Oppenheim}, where Tolman's law is assumed globally.}

\section{Universality of the Bekenstein-Hawking entropy 
}\lb{s:entropy}
We are now ready to derive the area law. 
In this local equilibrium, 
the 1D Gibbs relation 
\begin{equation}\lb{Gibbs}
T_{loc} s = \rho_{1d}+p_{1d}
\end{equation}
holds for $\rho_{1d}=4\pi r^2 \bra - T^{\hat t}{}_{\hat t}\ket$ and $p_{1d}=4\pi r^2 \bra T^{\hat r}{}_{\hat r}\ket$ 
because the interior is uniform in the proper radial length $\hat r$ \ci{Groot}.\footnote{Here, 
the chemical potential is zero because of the particle creation at each point inside (see Sec.\ref{s:T}) \ci{Landau_SM}.} 
Also, from \eqref{Trr}, 
\begin{equation}\lb{eos}
p_{1d}=\f{2-\eta}{\eta}\rho_{1d}
\end{equation}
plays a role of the equation of state, 
since $\bra T^\theta{}_\theta\ket (\gg \bra T^r{}_r\ket)$ 
has a qualitatively different origin (see Sec.\ref{s:self}). 
From \eqref{Ttt}, \eqref{L}, \eqref{T_loc}, \eqref{Gibbs} and \eqref{eos}, 
we obtain 
\begin{equation}\lb{s2}
s(r)=\f{1}{T_{loc}(r)}\f{2}{\eta}\rho_{1d}(r)=\f{2\pi \sqrt{2\s}}{l_p^2}.
\end{equation}
This provides the Bekenstein-Hawking entropy for any type of degrees of freedom: 
\begin{equation}\lb{S_th}
S=\int^R_0 dr \sqrt{g_{rr}(r)}s(r)=\int^R_0 dr \sqrt{\f{r^2}{2\s}} \f{2\pi \sqrt{2\s}}{l_p^2}
=\f{{\cal A}}{4l_p^2},
\end{equation}
where we have used \eqref{B} and ${\cal A}\equiv4\pi R^2\approx 4\pi a^2$. 
This saturates the entropy bounds \ci{B_bound,Bousso1}, even though there is no horizon. 

Let us discuss the reason for \eqref{S_th} by reviewing our self-consistent discussion. 
First, according to the concept of ``weight of information", 
we have considered the self-gravity of the quanta with 1-bit of information, 
made a \textit{rough mechanical estimation} of their entropy \eqref{S_eva}, 
and constructed $g_{rr}(r)$ \eqref{B} so that the entropy is \textit{proportional} to the area law. 
Here, the details of the degrees of freedom have been introduced into the metric as $\s$.\footnote{See \eqref{sigma} 
for the dependence of $\s$ on the type of fields.} 
We have also determined $g_{tt}(r)$ \eqref{A} from the acceleration condition \eqref{an}. 
Next, considering the obtained metric as the background,
we have shown that any degrees of freedom behave like the local thermal state at $T_{loc}(r)$, \eqref{T_loc}.  
Finally, using the Einstein equation \eqref{Einstein}, which will be shown to hold self-consistently in Sec.\ref{s:self}, 
we have evaluated the entropy density \eqref{s2} in an \textit{exact thermodynamical way}
and reached the area law \eqref{S_th}. 
Here, $\s$ cancels out exactly such that the precise coefficient $1/4$ appears universally for any interior degrees of freedom 
without causing a problem like the ``species problem" \ci{species1,species2}. 
This is a non-trivial result of the dynamics of gravity, \eqref{Einstein}. 

\section{Surface in thermal equilibrium}\lb{s:surface}
Let us fix the size $R$ and the constant $A_0$ from thermodynamics. 
Suppose that the configuration is in equilibrium with a heat bath of temperature $T$ (during time $\D t\ll \MO(e^{a^2/l_p^2})$). 
For simplicity, we neglect a small backreaction from thermal radiation 
and vacuum fluctuation \ci{BD} around the dense configuration 
and set the exterior metric for $r\geq R$ approximately as\footnote{
A (possibly-)large backreaction from quantum fluctuations near $r\sim a$ is 
considered in \eqref{interior0}. See Sec.\ref{s:self}.}  
\begin{equation}\lb{Sch}
ds^2=-\l(1-\f{a}{r}\r)dt^2+\l(1-\f{a}{r}\r)^{-1}dr^2+r^2 d\Omega^2.
\end{equation}
From the thermodynamic relation $TdS=dM(=d\frac{a}{2G})$ and \eqref{S_th}, 
then the equilibrium temperature is determined as 
\begin{equation}\lb{T_H}
T=\f{\hbar}{4\pi a},
\end{equation}
which agrees with the Hawking temperature. 
Furthermore, thermodynamics and Tolman's law\footnote{The redshift in \eqref{Sch} is so weak 
that radiations can propagate between the surface located at $r=R\approx a$ and the bath located at $r\gg a$, in a time, say, $\D t \sim \f{a^3}{l_p^2} (\ll \MO(e^{a^2/l_p^2}))$, 
and thus Tolman's law holds there.}
 require that the local temperature be continuous at $r=R$ \ci{Landau_SM}:   
\begin{equation}\lb{T_con}
T_{loc}(R)=\f{T}{\sqrt{1-\f{a}{R}}},
\end{equation}
where the left hand side is given by \eqref{T_loc} and $T$ by \eqref{T_H}. 
Setting $R=a+\D$ ($\D\ll a$), this becomes $\f{\hbar}{2\pi L}=\f{\hbar}{4\pi a}\f{1}{\sqrt{\f{\D}{R}}}\approx \f{\hbar}{4\pi \sqrt{a\D}}$, 
leading through \eqref{L} to $\D=\f{\s\eta^2}{2a}$. 
Thus, the position of the surface in thermodynamic equilibrium is determined as \eqref{R_I}:
\begin{equation}\lb{R}
R=a+\f{\s\eta^2}{2a}.
\end{equation}
We will see in Sec.\ref{s:T} that this position is consistent with Israel's junction condition \ci{Poisson}. 

To fix $A_0$, we use the continuity of the induced metric at $r=R$ \ci{Poisson}. 
This yields $-g_{tt}(R)=e^{\f{R^2}{2\eta\s}+A_0}=1-\f{a}{R}\approx \f{\s\eta^2}{2R^2}$, 
where \eqref{interior0}, \eqref{Sch}, and \eqref{R} have been used.  
Thus, the interior metric \eqref{interior0} is fixed completely as \eqref{interior_I}: 
\begin{equation}\lb{interior}
ds^2=-\f{\s\eta^2}{2R^2}e^{-\f{R^2-r^2}{2\s\eta}}dt^2+\f{r^2}{2\s}dr^2+r^2d\Omega^2.
\end{equation}

\eqref{R} indicates that the dense configuration has, instead of a horizon, 
the surface at $r=R>a$ as the boundary between \eqref{Sch} and \eqref{interior} \ci{KY2,KMY}.
Note that the proper length of the second term in \eqref{R}, 
$\D \hat r = \sqrt{g_{rr}(R)} \f{\s\eta^2}{2a}\sim \sqrt{\s}=\MO(\sqrt{N}l_p)$, is 
large compared to $l_p$ for $N\gg1$ and physically meaningful 
in the semi-classical description. 

\section{Self-consistent values of $(\s,\eta)$}\lb{s:self}
The results so far are based on the assumption that there exist $(\s,\eta)$ satisfying \eqref{Einstein}. 
To obtain them in a given theory, 
we need to evaluate the renormalized energy-momentum tensor $\bra \psi|T_{\mu\nu}|\psi\ket$ 
in the background metric \eqref{interior} and an excited state $|\psi\ket$ at $T_{loc}$, 
compare the both sides of \eqref{Einstein}, 
and then find the self-consistent values of $(\s,\eta)$. 
We here discuss this briefly. (See \ci{KY4} for more details.)

Because the energy scale inside is $\sim\f{m_p}{\sqrt{N}}$ (from \eqref{wave}, \eqref{RRR}, and \eqref{T_loc}) 
and close to the Planck scale, 
it should be reasonable to consider, for simplicity, 
the case where the degrees of freedom are conformal\footnote{Even for non-conformal fields, $\sigma$ can be determined \ci{KY4}.}. 
Then, we can use 4D Weyl anomaly:
\begin{equation}\lb{trace}
\bra \psi |T^\mu{}_\mu |\psi \ket=\hbar (c_W {\cal F}-a_W{\cal G}+b_W \Box R ),
\end{equation}
which is independent of $|\psi\ket$ \ci{BD}. 
Here, ${\mathcal F}\equiv C_{\a\b\g\d}C^{\a\b\g\d}=R_{\a\b\g\d}R^{\a\b\g\d}-2R_{\a\b}R^{\a\b}+\f{1}{3}\MR^2$ 
and 
${\mathcal G}\equiv R_{\a\b\g\d}R^{\a\b\g\d}-4R_{\a\b}R^{\a\b}+\MR^2$. 
$c_W$ and $a_W$ are positive constants determined by the matter content of the theory (for small coupling constants), 
while $b_W$ also depends on the finite renormalization of $R^2$ and $R_{\a\b}R^{\a\b}$ in the gravity action. 

Applying \eqref{RRR} and $\Box R=\MO(r^{-4})$ (for \eqref{interior}) to \eqref{trace}, 
the trace of \eqref{Einstein} becomes
\begin{equation}\lb{L2}
\f{2}{L^2}=8\pi l_p^2 c_W \f{4}{3L^4}\Rightarrow L^2=\f{16\pi l_p^2 c_W}{3}, 
\end{equation}
at the leading order for $r\gg l_p$. 
This leads through \eqref{L} to \ci{KY1,KY3}
\begin{equation}\lb{sigma}
\s=\f{8\pi l_p^2 c_W}{3\eta^2},
\end{equation}
where $c_W$ plays a role of $N$ in \eqref{B}. 
Noting that $c_W$ represents (roughly) the total number of the degrees of freedom in the theory \ci{BD}, 
we can expect that in general so does $N$. 
Therefore, for our description to be consistent, 
we need many kinds of fields so that $N=\MO(1)\gg1$, \eqref{large_N}. 

For $\eta$, we can use the dimensional regularization and a perturbative technique based on 
the warped product of $AdS_2$ and $S^2$ and evaluate directly another component, say, $\bra \psi|T^t{}_t|\psi\ket$,  
to find the self-consistent value of $\eta$ for the range \eqref{eta} (see \ci{KY4} for the details).

Then, the other components of \eqref{Einstein} hold automatically due to $\N_\mu \bra T^\mu{}_\nu\ket=0$. 
We thus conclude that the interior metric \eqref{interior} with \eqref{sigma} and \eqref{eta} 
satisfies the semi-classical Einstein equation \eqref{Einstein} self-consistently, 
and our assumptions, the uniformity and \eqref{an}, are consistent. 

The key in the proof above is a non-perturbative treatment of the full 4D dynamics. 
The 4D Weyl anomaly \eqref{trace} is a result of 
the vacuum fluctuation of all modes with arbitrary angular momentum \ci{BD,KY4}, 
which does not appear in 2D models or s-wave approximations. 
In \eqref{L2}, we have solved the trace part of \eqref{Einstein} non-perturbatively for $\hbar$ 
to obtain the self-consistent value of $L$. 
On the other hand, the acceleration \eqref{an}, pressure \eqref{T33}, and curvatures \eqref{RRR} 
can be characterized by $L\equiv\sqrt{2\eta^2\s}$, 
where $\hbar$ appear in their denominators, 
and thus we cannot take the limit $\hbar \to 0$. 
Therefore, the dense configuration \eqref{interior} is the non-perturbative solution for $\hbar$, 
where the full 4D dynamics is essential. 

\section{Dynamical derivation of $T_{loc}(r)$}\lb{s:T}
In Sec.\ref{s:thermal}, we have derived the local temperature \eqref{T_loc} 
based on the kinematical and universal argument with the Unruh effect. 
In this section, we rederive it from a direct discussion, 
which guarantees the local thermality dynamically and self-consistently. 
The argument consists of three steps. 
In step 1, we start with the results in Sec.\ref{s:metric} and rederive \eqref{T_H} 
as particle creation in the formation and evaporation process of the configuration; 
in step 2, we reobtain \eqref{R} from a mechanical equilibrium condition instead of thermodynamics; 
and in step 3, we consider an adiabatic formation of the configuration to reach \eqref{T_loc} dynamically.

\paragraph{\underline{Step 1.}} 
We study the time evolution of quantum matter fields 
in a background geometry for the formation and evaporation of the configuration, 
to show self-consistently that particles are created inside at temperature \eqref{T_H}. 

We first set up the background geometry. 
Suppose that the configuration with the interior metric \eqref{interior0} is formed from, say, a collapse of matter. 
The metric can be expressed as 
\begin{equation}\lb{interiorK}
ds^2=-K^2e^{-\f{R^2-r^2}{2\s\eta}}dt^2+\f{r^2}{2\s}dr^2+r^2d\Omega^2.
\end{equation}
At this stage, the radius $R$ of the surface is not yet determined except that $R=a+\D$ ($\D\ll a$). 
According to the continuity of the induced metric at $r=R$, 
a constant $K$ is determined by the exterior metric and the value of $\D$, 
such that $t$ connects to the static time in the asymptotically flat region. 
$K$ can then be a power of $a$: $K=K(a)$. 
As shown below, the details of $\D$ and $K$ are not important for the particle creation. 
For convenience, we now introduce a null time $u$ by 
$K e^{-\f{R^2}{4\s\eta}} du = K e^{-\f{R^2}{4\s\eta}}dt-\f{r}{\sqrt{2\s}}e^{-\f{r^2}{4\s\eta}}dr$, 
to rewrite \eqref{interiorK} as
\begin{equation}\lb{interiorK2}
ds^2=-K^2e^{-\f{R^2-r^2}{2\s\eta}}du^2-2K \f{r}{\sqrt{2\s}} e^{-\f{R^2-r^2}{4\s\eta}}dudr+r^2d\Omega^2,
\end{equation}
where $u$ is chosen to be connected to the outgoing null time in the asymptotically flat region.

Imagine then that the configuration evaporates in the vacuum according 
to the Stefan-Boltzmann law of the Hawking temperature: 
\begin{equation}\lb{SB_law}
\f{dM(u)}{du}=-\f{C}{2G a(u)^2}.
\end{equation}
Here, $M(u)=\f{a(u)}{2G}$ is the Bondi mass, 
and $C$ plays a role of the Stefan-Boltzmann constant of 
$\MO(N l_p^2)$\footnote{According to Sec.\ref{s:metric} and Sec.\ref{s:self}, 
$N$ corresponds to the number of the degrees of freedom contributing to the entropy 
and should also appear in the Stefan-Boltzmann constant \ci{Landau_SM}.} including the gray-body factor \ci{Landau_SM}. 
Then, the interior is almost frozen by the large redshift 
and is still described approximately by \eqref{interiorK2} 
with $R(u)=a(u)+\D'(u)~(\D'\ll a)$:\footnote{\lb{foot:DD}Since the evaporating configuration is not in equilibrium, 
its surface may not coincide with the surface of the stationary one in the heat bath; 
the difference may be $\D-\D'=\MO\l(Nl_p^2/a\r)$ due to the energy fluctuation $\sim \f{N\hbar}{a}$.}
\begin{equation}\lb{interior2}
ds^2=-K'^2e^{-\f{R(u)^2-r^2}{2\s\eta}}du^2
-2K' \f{r}{\sqrt{2\s}} e^{-\f{R(u)^2-r^2}{4\s\eta}}dudr+r^2d\Omega^2,
\end{equation}
where $K'$ can be a power of $a(u)$: $K'=K'(a(u))$.
On the other hand, the exterior is given by a $u$-dependent metric describing \eqref{SB_law}, say, a Vaidya metric \ci{Poisson}.  

Thus, the whole geometry for the formation and evaporation 
consists of the initial flat space, 
the intermediate curved space described by the above interior and exterior metrics, 
and the final flat space after a complete evaporation (which we assume). 
The Penrose diagram is topologically the same as the Minkowski space \ci{KMY,KY4,KY5}. 

Next, we suppose that in this time-dependent background spacetime, 
quantum fields start from the initial flat space in the Minkowski vacuum state $|0\ket_M$, 
pass the center $r=0$, propagate through the dense region with \eqref{interior2}, and come out to the future asymptotically flat region. 
Then, the fields feel the time-dependent gravitational potential to create particles \ci{Hawking,BD,Barcelo}. 
To demonstrate this simply, 
we consider s-waves of $N_s$ massless scalar fields in the eikonal approximation,
where the fields propagate along a $u$-constant line after passing $r=0$. 
We can then evaluate the energy flux observed at $r\gg a$ of the created particles as 
\begin{equation}\lb{J1}
J \equiv 4\pi r^2 {}_M\bra0|T_{uu}|0\ket_M|_{r\gg a} 
\approx \f{\hbar N_s}{48\pi} \l( \l(\f{d\xi}{du}\r)^2-2\f{d^2\xi}{du^2}\r), 
\end{equation}
where $\xi\equiv \log \f{du'}{du}$ and 
$u'$ is an arbitrary local time coordinate 
at a point $r=r'$ $(l_p\ll r'\ll R(u))$ 
on the $u$-constant line \ci{KMY,KY5} (see Appendix \ref{A:J} for a review). 

To evaluate \eqref{J1}, we can use, say, the proper time at $r=r'$ defined by 
$du'\equiv K' e^{-\f{R(u)^2-r'^2}{4\s\eta}}du $ (from \eqref{interior2}). 
Then, we have $\xi \approx -\f{R(u)^2-r'^2}{4\s\eta}$ and calculate 
$\f{d\xi}{du}\approx -\f{a(u)}{2\s\eta} \f{da(u)}{du}=\f{C}{2\s\eta a(u)}$ 
from $R(u)\approx a(u)$ and the ansatz \eqref{SB_law}, to reach 
\begin{equation}\lb{J2}
J=\f{\hbar N_s}{192\pi a(u)^2}\l(\f{C}{\s\eta}\r)^2.
\end{equation}
Comparing this to $J=\f{C}{2Ga^2}$ (from \eqref{SB_law}), 
we obtain 
\begin{equation}\lb{C}
C=\f{96\pi}{N_s l_p^2}\s^2\eta^2.
\end{equation}
This is $\MO(N_sl_p^2)$ indeed.
For example, 
applying \eqref{sigma} to $N_s$ scalar fields 
and using $c_W=\f{N_s}{1920\pi^2}$ \ci{BD}, 
\eqref{C} becomes $C=\f{N_s l_p^2}{5400 \pi \eta^2}$. 

Thus, the existence of $C$ proves self-consistently that 
particles are created inside as thermal-like radiation at the Hawking temperature \eqref{T_H} (more precisely, $T_H=\f{\hbar}{4\pi a(u)}$). 
Note that we can use a simple model to show this more explicitly \ci{KMY,KY5} (see also Appendix \ref{A:M_model}), 
and that we can also derive the particle creation inside at a 4D dynamics level \ci{KY3}.

\paragraph{\underline{Step 2.}}
Next, we put this evaporating configuration into a heat bath of temperature \eqref{T_H}. 
Then, the outgoing energy flow emitted from the configuration and the ingoing one from the bath are balanced, 
and the system reaches equilibrium. 
The exterior is given approximately by the Schwarzschild metric \eqref{Sch}, 
while the interior is described by \eqref{interior0} (or \eqref{interiorK}). 

To determine the position $R$ of the surface in a mechanical manner, 
we consider the surface energy-momentum tensor $S_{\mu\nu}$ on the timelike hypersurface at $r=R$, 
which represents, if any, a gap of energy-momentum density between the inside and outside geometries \ci{Poisson}. 
For mechanical equilibrium, it should be natural to assume that the surface is determined so that the gap is as small as possible 
in a consistent way with the uniform interior distribution. 
Note here that $N$ quanta with $\e_{loc}\sim\f{m_p}{\sqrt{N}}$ \eqref{wave} exist around each radius $r$ inside. 
Therefore, the condition should be given by 
\begin{equation}\lb{mech_con}
4\pi R^2 S^\mu{}_\nu\sim N\e_{loc}\sim \sqrt{N}m_p,
\end{equation}
which means that the quanta composing the surface at $r=R$ are 
responsible for the energy-momentum tensor $S_{\mu\nu}$ consistent with Israel's junction condition. 
Indeed, \eqref{mech_con} determines $R$ as \eqref{R}, leading to $S_{\mu\nu}$ that satisfies a stability condition for the surface \ci{Yang} (see Appendix \ref{A:surface}). 

Having obtained \eqref{T_H} and \eqref{R}, 
we can now fix the local temperature at $r=R$ from \eqref{T_con} as \eqref{T_loc}. 
Thus, the mechanical condition \eqref{mech_con} is consistent with the thermodynamical one \eqref{T_con}.

\paragraph{\underline{Step 3.}}
Finally, suppose that we grow up such a small configuration  
to a large one adiabatically by changing the temperature and size of the bath slowly \ci{KY1,KY5} 
(see Appendix \ref{A:M_model} for a simple model). 
The results of Step1 and Step2 hold at each stage of the formation because of the uniformity: 
at each stage with energy $\f{a'}{2G}$, 
the configuration creating particles at temperature $T=\f{\hbar}{4\pi a'}$ 
is in equilibrium with the bath of $T=\f{\hbar}{4\pi a'}$, 
the surface is located at $r=a'+\f{\s\eta^2}{2a'}$, 
and the local temperature there is \eqref{T_loc}. 
This local temperature is then kept in the subsequent stages due to the large redshift (during $\D t\ll \MO(e^{a^2/l_p^2})$). 

\paragraph{} Thus, we have shown that, 
as a result of the particle creation inside, 
the quanta in each small region behave like the local thermal state at the local temperature \eqref{T_loc}.
It is remarkable that the dynamical derivation here and the kinematic derivation of Sec.\ref{s:thermal} give the same result, \eqref{T_loc}.

\section{Conclusion and discussion}\lb{s:con}
In this paper, we have constructed self-consistently the semi-classical configuration consisting of many excited quanta with the strong self-gravity and large quantum effects (as in Fig.\ref{f:Whole}):
the interior metric represents the dense compact object with the near-Planckian curvature and satisfies non-perturbatively the 4D semi-classical Einstein equation with many matter fields ($N\gg1$), and the entropy follows the Bekenstein-Hawking formula universally for any local degrees of freedom. 
In this sense, this configuration can be a candidate for quantum black holes. We thus conclude that the strong self-gravity plays the key role in changing the entropy from the volume law to the area law. We here write the interior metric \eqref{interior} again: 
\begin{equation}
ds^2=-\f{\s\eta^2}{2R^2}e^{-\f{R^2-r^2}{2\s\eta}}dt^2+\f{r^2}{2\s}dr^2+r^2d\Omega^2 ~~~~{\rm for}~~\sqrt{N} l_p\lesssim r \leq R.\tag{\ref{interior}}
\end{equation}

It is remarkable 
that the dense configuration is robust in that the metric \eqref{interior} can be obtained by various approaches \ci{KMY,KY2,KY1,Ho,KY3,KY4,KY5}. 
Compared to these previous studies, the novelties of this paper are as follows. 
(i) We have directly considered the effect of self-gravity on entropy and constructed the interior metric from the proportionality of the area law. 
This method clearly shows the importance of the self-gravity in the entropy-area law. 
(ii) Using the Unruh temperature and thermodynamics locally, the Bekenstein-Hawking formula has been obtained exactly for any value of $\eta$ and any type of the degrees of freedom. 
This shows the universality of the area law in this configuration. 
(iii) The surface location and local temperature have been determined by two methods: thermodynamic and mechanical equilibrium conditions. 
This agreement makes the thermal behavior more robust. 
In summary, the present study provides results that directly fuse quantum theory, gravity, and thermodynamics at a semi-classical level.

Now, we compare our results/approaches to other horizon/singularity-avoiding scenarios. 
As a semi-classical one, Ref.\ci{Terno} constructs a metric around a regular apparent horizon without any drastically large energy-momentum tensor, study the motion of a thin shell in the metric by using a 2D junction condition not considering \eqref{Einstein} self-consistently, to show that the shell collapses to form the horizon. 
On the other hand, the multi-shell model in Ref.\ci{KMY} (and Appendix \ref{A:M_model}) considers both the time evolution of continuously many null shells and the backreaction of the evaporation in a common time coordinate and examines whether an apparent horizon is formed, without assuming anything about the magnitude of the energy-momentum tensor a priori. 
Solving \eqref{Einstein} self-consistently, each shell will never cross its shrinking Schwarzschild radius, and a surface pressure will occur on each shell \ci{KMY}, which becomes the large tangential pressure \eqref{T33} in a continuum limit, leading to the dense configuration without forming horizons. Thus, the 4D dynamics of \eqref{Einstein} including the tangential direction makes the difference. 

As a quantum-gravity approach, Ref.\ci{Edward} considers a spherically-symmetric quantum-gravity model and analyzes the time evolution of a collapsing matter to provide a bounce scenario in which a horizon is formed but no singularity appears due to a repulsive effect induced non-perturbatively. Note here that the 4D dynamics of quantum matter fields is not considered in the analysis. In our scenario, as explained just above, the large tangential pressure plays a key role in supporting the dense configuration without horizons or singularities, and it is a non-perturbative effect coming from vacuum fluctuations of all modes with an angular momentum \ci{KY4} (see Sec.\ref{s:self} again). It would be interesting to use an idea of quantum-gravity dynamics as in Ref.\ci{Edward} to describe the regions beyond the semi-classical approximation (see also below). 

Next, let us consider the limitations 
of our approach to quantum black holes. First, our arguments have started from the two simple conditions (the uniformity and maximum acceleration), 
which can be considered as sufficient conditions for the Bekenstein-Hawking entropy. 
To better understand the validity of the two assumptions, it would be important to consider the necessary conditions for the area law in a more general framework. 
One expectation can come from thermodynamics. 
As shown in Appendix \ref{A:M_model}, 
these conditions are naturally satisfied in the adiabatic formation process 
and should be rather typical, since adiabatic processes generally lead to most typical configurations in thermodynamics \ci{Landau_SM}. 

Second, the metric \eqref{interior} and the Bekenstein-Hawking formula \eqref{S_th} should hold for a broad class of local quantum fields. In Sec.\ref{s:self}, we have used, for simplicity, conformal matter fields. 
However, the energy scale $\sim m_p/\sqrt{N}$ is close to the Planck scale, and we can expect that the metric \eqref{interior} and the area law \eqref{S_th}  are valid for any local fields 
as long as the number of the degrees of freedom $N$ is large and mass of a field is smaller than $\sim m_p/\sqrt{N}$, since such a mass would be negligible as a leading effect. 
Of course, the values of $(\sigma,\eta)$ depend on the type of the fields. 
We can check explicitly these expectations for non-conformal massless scalar fields \ci{KY4}. 

Third, our semi-classical approximation based on \eqref{Einstein} cannot be applied to the small center region and the last stage of the evaporation. As discussed below \eqref{L}, the size of the small center region is $\sim l_p$, which is smaller than the finest resolution in a semi-classical spacetime, 
and the region cannot be described by \eqref{Einstein}. 
Similarly, the approximation would break down at the last moment of the evaporation in the model in Appendix \ref{A:M_model}, 
since extrapolating the Stefan-Boltzmann law \eqref{model2} to $a_i\to0$ would exceed the Planck energy.
To describe such regions, therefore, we would need ideas beyond the present approximation. 

Finally, let's consider two future prospects inspired by our picture of the entropy-area law. 

The first one is an implication to black holes in quantum gravity. As an interesting proposal, Ref.\ci{QG_con} uses a hydrodynamic approximation of group field theory (a second quantization version of loop quantum gravity) \cite{Oriti}, which inherently has no notion of continuum spacetime geometry, and constructs a quantum gravity state for spherically symmetric geometries, to represent a black hole as a quantum gravity condensate. Similarly to our approach, they characterize the black hole not by a geometric horizon condition but by an entropy condition (its maximization).

More specifically, they construct a spherically symmetric condensate state by ``gluing" many spherical shell states that represent quantum-gravity quanta, whose picture is similar to Fig.\ref{f:Whole}. In our case, the width of the region in which time effectively elapses is $\Delta r\sim \sqrt{\sigma}/r$ ($\sim \sqrt{N}l_p$ as the proper length) due to the large redshift in the metric \eqref{interior}, and we can consider that the spherically excited quanta of width $\sim \sqrt{N}l_p$ gather in a continuous concentric way to form the dense configuration as in Fig.\ref{f:Whole}. Here, an interesting common feature is that the quanta live in the whole bulk but the entropy is proportional to the surface area. In their case, this holographic behavior comes from a property of the condensate state and holds for any size. In our case, too, the area law \eqref{S_th} holds at any radius $r$, which is a consequence of the uniformity. In this sense, 
both exhibit universal behavior with respect to the radial direction and are similar. 

Their result appears by taking a kind of semi-classical limit in the non-perturbative formulation of quantum gravity (without matter degrees of freedom), while our result is obtained by solving the semi-classical Einstein equations (with many matter fields) in a non-perturbative way. Therefore, the above similarity may tell us that these two different descriptions are somehow related. Let us here observe the difference in the treatment of dynamics. Their condensate state is chosen by a kinematical argument without considering the equation of motion, while our state and geometry are determined dynamically by solving the semi-classical Einstein equation \eqref{Einstein}. This point leads to the difference in how the Bekenstein-Hawking entropy is obtained.  In their calculation, applying the entropy maximization condition makes the entanglement entropy for the outermost shell proportional to the surface area, and the proportionality coefficient is chosen to fit the Bekenstein-Hawking formula as a thermodynamic consistency of the whole system.  In our case, on the other hand, the Unruh effect is applied locally to the self-consistent geometry \eqref{interior} to get the local temperature \eqref{T_loc} (whose origin can be understood dynamically as in Sec.\ref{s:T}); and then, by using local thermodynamics in subsystems, the entropy density \eqref{s2} is evaluated and summed over the bulk interior to reach the area law including the factor $1/4l_p^2$. 

These suggest that applying such local thermodynamic consistency of a subsystem, rather than thermodynamic consistency of the whole system, may be a more appropriate way to deal with the dynamics. Considering the entropy density and local temperature in a quantum-gravity formulation and imposing a behavior similar to \eqref{T_loc} and \eqref{s2} may provide a more appropriate description of black holes in quantum gravity and lead to a more natural derivation of the Bekenstein-Hawking entropy. We would like to explore this idea in the future.

The second prospect is a view of entropy including self-gravity as a gravitational charge. 
The only component of the semi-classical Einstein equation \eqref{Einstein} used for the mechanical estimation \eqref{s_def} of the entropy density $s(r)$ \footnote{For convenience, we recall \eqref{s_def} here: 
\begin{equation}
s(r)\sim \f{N}{\D \hat r}\sim \f{N\sqrt{g_{rr}(r)}}{r},\tag{\ref{s_def}}
\end{equation}
where the second estimation is the result of \eqref{ADM}.} 
is essentially the quasi-local mass formula \eqref{ADM}: 
\begin{equation}
M(r)=4\pi \int^r_0 dr' r'^2 \bra-T^t{}_t(r')\ket.  \tag{\ref{ADM}}
\end{equation}
Of course, the final exact form \eqref{S_th} is guaranteed by self-consistently satisfying all the components of the Einstein equation. 
However, \eqref{ADM} should play a special role in taking into account the effect of self-gravity on entropy. 

We note here that \eqref{ADM} is a result of the Hamiltonian constraint. 
In the general static spherically-symmetric metric \eqref{metric}, 
setting $g_{rr}(r)=\l(1-\f{a(r)}{r}\r)^{-1}$ for $a(r)\equiv 2G M(r)$, 
the Hamiltonian constraint $\MH=0~(\Leftrightarrow G_{tt}=8\pi G \bra T_{tt}\ket)$ is given by 
\begin{equation}
\p_r a(r)=8\pi G r^2 \bra -T^t{}_t\ket, 
\end{equation}
which is integrated to provide \eqref{ADM}.  
Indeed, the Hamiltonian constraint is important in obtaining the entropy $S$ of self-gravitating thermal radiation \ci{Sorkin}. 
On the other hand, a gravitational charge $Q_\xi$ (Noether/Hamiltonian charge) of a bounded subsystem 
associated with a diffeomorphism $\xi^\mu$, generally, 
consists of the sum of a bulk integral, which vanishes when the Hamiltonian and momentum constraints are satisfied, 
and a surface integral that can take nontrivial values depending on the nature of $\xi^\mu$ and the boundary \ci{Compere}. 
Therefore, the on-shell charge is given by the surface integral, which is similar to the above cases of entropy with self-gravity. 

There is another remarkable point. 
Black hole entropy can be formulated as a Noether charge associated with 
the symmetry for an infinitesimal time translation $v\to v+\e \hbar \b_H$, 
where $v$ is the Killing parameter and $\b_H^{-1}$ is the Hawking temperature \ci{Wald}. 
On the other hand, the entropy of matter without self-gravity is generally conserved under adiabatic quasi-static processes (adiabatic invariance) \ci{Landau_SM}, 
and it can be formulated as a Noether invariant associated with the symmetry for $t\to t+\e \hbar \b(t)$, 
where $t$ is a time coordinate and $\b(t)^{-1}$ is the temperature at time $t$ \ci{SY,SSY}. 
The two symmetries coincide, leading to a possibility that such a symmetry ``unifies" 
matter entropy and gravity entropy. 

From these discussions, 
we can expect the following. 
\textit{The entropy of a gravitational system is given ``holographically" 
by the gravitational charge evaluated on its boundary surface associated with a certain symmetry, representing matter entropy and gravity entropy in a unified manner. In a strong gravity limit, the charge becomes the Bekenstein-Hawking formula, and such a subsystem is considered as a black hole.}

In our picture, such a black hole is given by the dense configuration as in Fig.\ref{f:Whole}, and the interior structure provides a possible way to recover information consistently with a microscopic view of this unified entropy. As seen in Sec.\ref{s:T} and Appendix \ref{A:M_model}, particle creation occurs at each point inside, where the excited modes exist that represent the collapsing matter with the physical information $|\psi\ket$. The local energy scale, given by \eqref{T_loc}, is close to the Planck scale, and some quantum-gravity effect should appear approximately and induce an interaction between the outgoing created particles and the incoming matter \cite{HKY}. 
Indeed, we can use a simple scattering model and estimate the scattering time scale as $\Delta t\sim a\log \frac{a}{\sqrt{N}l_p \lambda}$, where $\lambda$ is a dimensionless coupling constant \cite{KY2}.
Through many such interactions, the particles emitted during the evaporation may extract the information out together with the energy \ci{KY4}, reproducing the Page curve \cite{Page}. We leave a detailed analysis of this speculation for the future. 


\section*{ACKNOWLEDGMENTS}
Y.Y. thanks K.Goto, P.H\"{o}hn, S.Nagataki, R.Norte, and N.Oshita for valuable comments and encouragement.
Y.Y. is partially supported by Japan Society of Promotion of Science 
(Grants No.21K13929 and No.18K13550) and by RIKEN iTHEMS Program. 
\appendix

\section{Multi-shell model}\lb{A:M_model}
For a self-contained argument, 
we provide a short review of a simple model that describes adiabatic formation of 
a uniform configuration in a heat bath \ci{KMY,KY4}. 
Specifically, we analyze the time evolution of a null matter representing radiation coming from a heat bath at the Hawking temperature, 
including the backreaction of particles created during the collapse, 
to show explicitly that the matter approaches a uniform configuration and eventually evaporates, 
and that the emitted particles behave like a thermal state at the Hawking temperature. 
Note that this model is just an example of the configuration in the main text. 

\begin{figure}[h]
\begin{center}
\includegraphics*[scale=0.12]{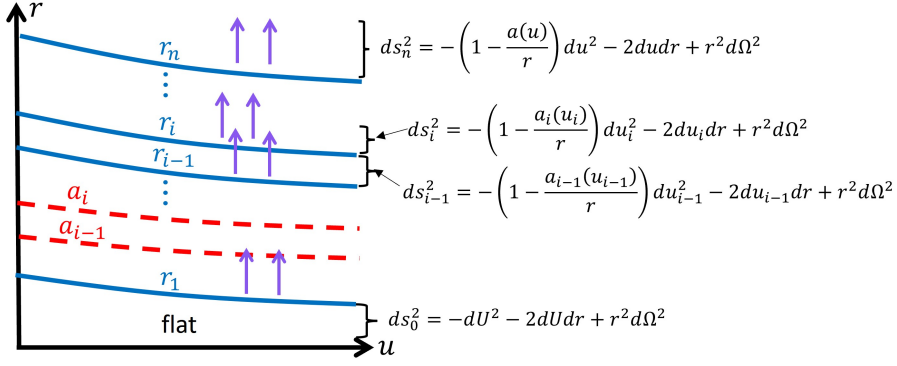}
\caption{\footnotesize A multi-shell model. 
$n$ spherical null shells (radiations from a heat bath) come together, creating and emitting particles dynamically.}
\label{f:multi}
\end{center}
\end{figure}

Suppose that, during adiabatic formation in a heat bath, 
many radiations with typical energy $\sim\f{\hbar}{a}$ come together in order with spherical symmetry. 
We model these radiations as $n(\gg1)$ concentric thin null shells. 
Then, the metric changes in time, and generically particle creation can occur around each shell \ci{BD,Hawking,Barcelo}. 
To introduce its backreaction, 
we postulate, as an ansatz, that each shell evaporates according to 
the Stefan-Boltzmann law of Hawking temperature, \eqref{SB_law}, 
and that the metric just outside each shell is given by the Vaidya metric \ci{Poisson}:
\begin{equation}\lb{model1}
ds^2_i=-\l(1-\f{a_i(u_i)}{r}\r)du_i^2-2du_idr+r^2 d\Omega^2,
\end{equation}
with
\begin{equation}\lb{model2}
\f{da_i(u_i)}{du_i}=-\f{C}{a_i^2}
\end{equation}
for $i=0,1,2\cdots n$. 
See Fig.\ref{f:multi}. 
Here, 
$\f{\D a_i}{2G}\equiv \f{a_i-a_{i-1}}{2G}$ is the energy of the $i$-th shell, 
and $u_i$ is the local time just above it. 
The total size $a_n$ is $a$, the outer most time $u_n$ is $u$, and the center is flat: 
$u_n=u,~a_n=a;~u_0=U,~a_0=0.$
To connect these coordinates, we employ the fact that 
each shell, whose locus is denoted by $r=r_i(u_i)$, moves at the speed of light in both its exterior and interior metrics:
\begin{equation}\lb{model3}
\f{r_i-a_i}{r_i}du_i=-2dr_i=\f{r_i-a_{i-1}}{r_i}du_{i-1}.
\end{equation}
This means 
\begin{align}\lb{model4}
\f{dr_i(u_i)}{du_i}&=-\f{r_i(u_i)-a_i(u_i)}{2r_i(u_i)},\\
\lb{model5}
\f{du_i}{du_{i-1}}&=\f{r_i-a_{i-1}}{r_i-a_i}=1+\f{a_i-a_{i-1}}{r_i-a_i}.
\end{align}

We here examine where $r_i(u_i)$ will approach in the background \eqref{model1}. 
When $r_i\sim a_i$, we can replace $r_i$ in the denominator of \eqref{model4} by $a_i$ 
and set $\D r_i(u_i)\equiv r_i(u_i)-a_i(u_i)$ to obtain 
\begin{equation}\lb{model6}
\f{d \D r_i(u_i)}{d u_i}\approx -\f{\D r_i(u_i)}{2a_i(u_i)}-\f{da_i(u_i)}{du_i}. 
\end{equation}
The first term is negative, expressing the effect of the collapse, 
and the second one is positive due to the evaporation \eqref{model2}. 
When $\D r_i(u_i)\sim \f{Nl_p^2}{a_i(u_i)}$, 
both terms are balanced and the right-hand side approaches zero with time. 
Therefore, each shell will behave asymptotically as \ci{KY2}
\begin{equation}\lb{model7}
r_i(u_i)\to a_i(u_i)-2a_i(u_i) \f{da_i(u_i)}{da_i}= a_i(u_i)+\f{2C}{a_i(u_i)}.
\end{equation}

Now, let us take a continuum limit $\D a_i\to 0$ and assume that each shell reaches the asymptotic position \eqref{model7}. 
Then, the shells pile up continuously and become a spherical dense configuration with the total mass $\f{a}{2G}$, as in Fig.\ref{f:Whole}. 
We can get the redshift factor $\xi_i\equiv\log \f{dU}{du_i}$, as follows. 
\begin{align}
\xi_i-\xi_{i-1} &=\log \f{\f{dU}{du_i}}{\f{dU}{du_{i-1}}}=-\log\f{du_i}{du_{i-1}} \nn\\
 &=-\log\l(1+\f{a_i-a_{i-1}}{r_i-a_i}\r) \nonumber \\
 &\approx- \f{a_i-a_{i-1}}{r_i-a_i} = -\f{a_i-a_{i-1}}{\f{2C}{a_i}} \nonumber \\
 &\approx -\f{1}{4C} (a_i^2-a_{i-1}^2).
\end{align}
Here, at the second line we used \eqref{model5}; 
at the third line we used $C\sim Nl_p^2$ and considered $\f{a_i-a_{i-1}}{\f{2C}{a_i}}\sim \f{l_p^2/a_i}{Nl_p^2/a_i}=\f{1}{N}\ll1$ 
for a large $N$ (or $\D a_i \to 0$); 
at the final line we approximated $2a_i \approx a_i+a_{i-1}$. 
Using $u_0=U$ and $a_0=0$, 
we obtain
\begin{equation}\lb{model8}
\xi_i =-\f{1}{4C}a_i^2.
\end{equation}

Then, we can construct the continuum interior metric in the $(U,r)$ coordinate. 
By considering the shell that passes a spacetime point $(U,r)$ inside, 
we have at $r=r_i$ 
\begin{equation}
\f{r_i-a_i}{r_i}=\f{\f{2C}{a_i}}{r_i}\approx \f{2C}{r^2},~~
\f{du_i}{dU} =e^{-\xi_i}=e^{\f{a_i^2}{4C}}\approx e^{\f{r^2}{4C}}, \nn
\end{equation}
where we employed \eqref{model7} and \eqref{model8}. 
Thus, the metric at that point is given by  
\begin{align}\lb{model9-1}
ds^2&=-\l(1-\f{a_i}{r_i}\r)du_i^2-2du_idr+r^2_i d\Omega^2,\nn\\
 &=-\l(1-\f{a_i}{r_i}\r)\l(\f{du_i}{dU}\r)^2dU^2-2\l(\f{du_i}{dU}\r)dUdr+r^2_i d\Omega^2 \nonumber \\
 &\approx -\f{2C}{r^2} e^{\f{r^2}{2C}}dU^2-2 e^{\f{r^2}{4C}}dUdr+r^2d\Omega^2.
\end{align} 

By introducing $e^k dt=dU+\f{r^2}{2C}e^{-\f{r^2}{4C}}dr$ ($k$: a constant or a function of $t$), 
\eqref{model9-1} can be rewritten as 
\begin{equation}\lb{model9}
ds^2 =-\f{2C}{r^2}e^{\f{r^2}{2C}+2k}dt^2+\f{r^2}{2C}dr^2+r^2d\Omega^2.
\end{equation}
Comparing this with \eqref{interior}, 
the power parts of $r$ in $g_{tt}(r)$ are different 
but don't contribute to the leading values (for $r\gg l_p$) of the curvatures and accelerations. 
Therefore, the both are the same (at the leading level) when we identify 
\begin{equation}\lb{model10}
\s=C,~~\eta=1
\end{equation}
and choose $k$ properly. 
Note that the position of the surface of the evaporating configuration in this model is 
given from \eqref{model7} by $R(u)=a(u)+\f{2C}{a(u)}$, 
which may be different from that of 
the stationary configuration in the heat bath (see footnote \ref{foot:DD}).

Finally, we consider s-waves of $N_s$ massless scalar fields in the eikonal approximation 
and apply the formula of the energy flux \eqref{J2} 
to \eqref{model10}, to obtain 
\begin{equation}\lb{model11}
J=\f{\hbar N_s}{192\pi a(u)^2}.
\end{equation}
Using this, \eqref{model10} and $J=\f{C}{2G a(u)^2}$ (from \eqref{model2}), 
we can find the self-consistent value of $\s~(=C)$ in this model: 
\begin{equation}\lb{model12}
\s_{model}=\f{N_s l_p^2}{96\pi}.
\end{equation}
Here, $\eta=1$ leads to $\bra T^r{}_r\ket=-\bra T^t{}_t\ket$ (from \eqref{Trr}), 
which means that the radial motion of the quanta is null without scattering \ci{KY1}. 
Considering this point and the spherical symmetry of the system,  
we can identify \eqref{model11} with the energy flux of 
1D thermal radiation $J_{1d}=\f{N_s}{12\hbar}T^2$ \ci{Landau_SM}, 
to get the Hawking temperature:
\begin{equation}\lb{model13}
T=\f{\hbar}{4\pi a(u)}.
\end{equation}
Note that the above analysis can be applied to each $a_i$. 

Thus, we have seen explicitly that 
(i) as a result of the self-consistent time evolution of the adiabatic formation, 
the uniform dense configuration appears 
(see \ci{KY4} for a more direct analysis of the semi-classical time evolution), 
and (ii) particles are created at temperature $T=\f{\hbar}{4\pi a_i}$ in each region inside 
(see \ci{KMY} for a Planck-like spectrum).


\section{Derivation of energy flux formula $J$}\lb{A:J}
We give a short review of the derivation of \eqref{J1} \ci{KY5}. 
First, we consider the s-waves of the $N_s$ scalar fields propagating in the time-dependent geometry 
described above \eqref{J1}. 
We can then use the eikonal solution and 
the point-splitting regularization, to express the energy flux of the particles created during the propagation as \ci{KMY}
\begin{equation}\lb{J_der1}
J \equiv 4\pi r^2 {}_M\bra0|T_{uu}|0\ket_M|_{r\gg a} = \f{\hbar N_s}{16\pi} \{u,U\}.
\end{equation}
Here, $\{x,y\} \equiv \f{\ddot y^2}{\dot y^2}-\f{2}{3}\f{\dddot y}{\dot y}$ 
is the Schwarzian derivative for $y=y(x)$, and 
$U$ is the outgoing null time in the flat space before the formation. 

Using a formula $\{z,x\}=\{z,y\}+\l(\f{dy}{dx}\r)^2\{y,x\}$, 
this can be expressed as 
\begin{equation}\lb{J_der2}
J= \f{N_s\hbar}{16\pi} \l(\{u,u'\}+\l(\f{du'}{du}\r)^2\{u',U\}\r),  
\end{equation}
where $u'$ is an arbitrary local time coordinate at $r=r'(<R(u))$ on the $u$-constant line. 
The first term represents the energy flux of the particles created in the region $r'\lesssim r \lesssim R(u)$, 
and the second one expresses the redshifted energy flux of those below $r=r'$. 
Therefore, if we take, say, the proper time at $r=r'$ defined by 
$du'\equiv K' e^{-\f{R(u)^2-r'^2}{4\s\eta}}du $ (from \eqref{interior2})
and consider $r'\ll R(u)-\f{\s}{a(u)}$, 
the second term is negligible, 
and only the first term remains. 
Then, using another expression $\{u,u'\}=\f{1}{3}\l(\f{d\xi}{du}\r)^2-\f{2}{3}\f{d^2\xi}{du^2}$ 
for $\xi\equiv \log \f{du'}{du}$, 
we reach 
$J\approx \f{\hbar N_s}{48\pi} \l(\l(\f{d\xi}{du}\r)^2-2\f{d^2 \xi}{du^2}\r)$, 
which is \eqref{J1}.

\section{Fixing the surface from mechanical equilibrium}\lb{A:surface}
We show that the mechanical equilibrium condition \eqref{mech_con} fixes $R$ as \eqref{R}.
First, the surface energy-momentum tensor on the timelike hypersurface $\Sigma$ at $r=R$ is given by \ci{Poisson} 
\begin{equation}\lb{SEMT}
S_{\mu\nu}=-\f{1}{8\pi G}([K_{\mu\nu}]-h_{\mu\nu}[K]).
\end{equation}
Here, we write the exterior and interior metrics collectively as the form of \eqref{metric}, 
and we define the induced metric
$h_{\mu\nu}dx^\mu dx^\nu\equiv -e^{A(R)}dt^2+R^2d\Omega^2$, 
the extrinsic curvature 
$K_{\mu\nu}\equiv h_\mu{}^\a h_\nu{}^\b \N_\a r_\b$ and its trace $K\equiv K^\mu{}_\mu$, 
and the ``jump" of a quantity $X$ across $\Sigma$ $[X]\equiv X_+-X_-$, 
where $r_\mu dx^\mu=\sqrt{B}dr$ is the unit normal vector to $\Sigma$ 
and $X_\pm\equiv \lim_{r\to R \pm0}X$ are $X$ on $\Sigma$ in the interior/exterior region. 

From these, we can calculate
$K^t{}_t=\a_n$ and $K^\theta{}_\theta=K^\phi{}_\phi=\f{1}{r\sqrt{B}}$, 
where $\a_n=\f{\p_r A}{2\sqrt{B}}$. Using the interior metric \eqref{interior0} and the exterior one \eqref{Sch}, 
we can check 
\begin{align}\lb{S1}
[\a_n] &=\f{\f{a}{R^2}}{2\sqrt{1-\f{a}{R}}}-\f{1}{\sqrt{2\s\eta^2}} \nn \\
 &\approx \f{1}{2\sqrt{a\D}}-\f{1}{\sqrt{2\s\eta^2}}=\MO\l(\f{1}{\sqrt{N}l_p}\r),\\
\lb{S2}
\l[\f{1}{r\sqrt{B}}\r]&=\f{1}{R}\sqrt{1-\f{a}{R}}-\f{\sqrt{2\s}}{R^2}\nn\\
&\approx\f{1}{a^2}(\sqrt{a\D}-\sqrt{2\s})=\MO\l(\f{\sqrt{N}l_p}{a^2}\r),
\end{align}
where $R=a+\D$ and $\D\ll a$. 
Therefore, \eqref{mech_con} together with \eqref{SEMT} requires $[\a_n] =0$. 
This and \eqref{S1} lead to $\D=\f{\s\eta^2}{2a}$, which means \eqref{R}. 

As a result, the surface energy density and pressure are given by, respectively, 
\begin{equation}
-S^t{}_t=\f{1}{4\pi R^2} \f{2-\eta}{G}\sqrt{\f{\s}{2}},~~
S^\theta{}_\theta= -\f{1}{4\pi R^2} \f{2-\eta}{2G}\sqrt{\f{\s}{2}}. 
\end{equation}
For \eqref{eta}, the former is positive and the latter is negative, 
which satisfies a dynamical stability condition for the surface \ci{Yang}.



\begin{thebibliography}{99}


\bibitem{Landau_SM}
L. D. Landau and E. M. Lifshitz, \textit{Statistical Physics} (Butterworth-Heinemann, Oxford, 1984).


\bibitem{Gross}
D.~J.~Gross, R.~D.~Pisarski and L.~G.~Yaffe,
Rev. Mod. Phys. \textbf{53}, 43 (1981). 

\bibitem{Pad_thermo}
T.~Padmanabhan,
Phys. Rept. \textbf{188}, 285 (1990).



\bibitem{Lynden}
D.~Lynden-Bell and R.~Wood,
Mon. Not. Roy. Astron. Soc. \textbf{138}, 495 (1968).

\bibitem{Tolman}
R.~Tolman and P.~Ehrenfest,
Phys. Rev. \textbf{36}, no.12, 1791-1798 (1930).

\bibitem{Sorkin}
R.~D.~Sorkin, R.~M.~Wald and Z.~J.~Zhang,
Gen. Rel. Grav. \textbf{13}, 1127-1146 (1981). 

\bibitem{Bekenstein} 
J.~D.~Bekenstein,
Phys.\ Rev.\ D {\bf 7}, 2333 (1973); Phys.\ Rev.\ D {\bf 9}, 3292 (1974).

\bibitem{Hawking} 
  S.~W.~Hawking,
  Commun.\ Math.\ Phys.\  {\bf 43}, 199 (1975)  [Erratum-ibid.\  {\bf 46}, 206 (1976)].  


\bibitem{B_bound}
J.~D.~Bekenstein,
Phys. Rev. D \textbf{23}, 287 (1981).

\bibitem{Bousso1}
R.~Bousso,
JHEP \textbf{07}, 004 (1999).

\bibitem{tHooft}
G.~'t Hooft,
Conf. Proc. C \textbf{930308} (1993), 284-296. 

\bibitem{Susskind}
L.~Susskind,
J. Math. Phys. \textbf{36} (1995), 6377-6396. 

\bibitem{Bousso2}
R.~Bousso,
Rev. Mod. Phys. \textbf{74}, 825-874 (2002). 

\bibitem{Oppenheim}
J.~Oppenheim,
Phys. Rev. D \textbf{65}, 024020 (2002); Phys. Rev. E \textbf{68}, 016108 (2003). 

\bibitem{Pesci}
A.~Pesci,
Class. Quant. Grav. \textbf{24}, 2283-2300 (2007). 

\bibitem{Poisson}
E. Poisson, \textit{A Relativistic Toolkit} (Cambridge University Press, Cambridge, 2004).

\bibitem{Pad}
T.~Padmanabhan,
Class. Quant. Grav. \textbf{4}, L107-L113 (1987)

\bibitem{Yoneya}
T.~Yoneya,
Prog. Theor. Phys. \textbf{103}, 1081-1125 (2000).

\bibitem{string}
J.~Polchinski, \textit{String Theory} (Cambridge University Press, Cambridge, 2005). 

\bibitem{Loop}
C.~Rovelli and F.~Vidotto, \textit{Covariant Loop Quantum Gravity} 
(Cambridge University Press, Cambridge, 2014). 

\bibitem{Cai1}
E.~R.~Caianiello,
Lett. Nuovo Cim. \textbf{32}, 65 (1981)


\bibitem{Brandt}
H.~E.~Brandt,
Found. Phys. Lett. \textbf{2} (1989), 39.




\bibitem{RV}
C.~Rovelli and F.~Vidotto,
Phys. Rev. Lett. \textbf{111}, 091303 (2013).

\bibitem{Buchdahl}
H.~A.~Buchdahl,
Phys. Rev. \textbf{116}, 1027 (1959).


\bibitem{KMY} 
H.~Kawai, Y.~Matsuo, and Y.~Yokokura,
Int.\ J.\ Mod.\ Phys.\ A {\bf 28}, 1350050 (2013).  

\bibitem{KY1}
  H.~Kawai and Y.~Yokokura, 
  Int.\ J.\ Mod.\ Phys.\ A {\bf 30}, 1550091 (2015).  
 
\bibitem{KY2} 
  H.~Kawai and Y.~Yokokura,
  Phys.\ Rev.\ D {\bf 93}, no. 4, 044011 (2016). 

\bibitem{Ho}
P.~M.~Ho,
Class. Quant. Grav. \textbf{34}, no.8, 085006 (2017).

\bibitem{KY3}
H.~Kawai and Y.~Yokokura,
Universe \textbf{3}, no.2, 51 (2017).

\bibitem{KY4}
H.~Kawai and Y.~Yokokura,
Universe \textbf{6}, no.6, 77 (2020). 

\bibitem{KY5}
H.~Kawai and Y.~Yokokura,
Phys. Rev. D \textbf{105}, no.4, 045017 (2022).

\bibitem{Landau_C}
L. D. Landau and E. M. Lifshitz, \textit{The Classical Theory of Fields} (Butterworth-Heinemann, Oxford, 1980).

\bibitem{BD}
N.~D.~Birrell and P.~C.~W.~Davies, \textit{Quantum Fields in Curved space} (Cambridge University Press, Cambridge, 1982). 

\bibitem{Unruh}
W.~G.~Unruh,
Phys. Rev. D \textbf{14}, 870 (1976)

\bibitem{Jacobson}
T.~Jacobson,
Phys. Rev. Lett. \textbf{75}, 1260-1263 (1995).

\bibitem{Pad2}
T.~Padmanabhan,
Rept. Prog. Phys. \textbf{73}, 046901 (2010).

\bibitem{Caianiello}
E.~R.~Caianiello and G.~Landi,
Lett. Nuovo Cim. \textbf{42}, 70 (1985)


\bibitem{Groot}
S. R. de Groot and P. Mazur, \textit{Non-Equilibrium Thermodynamics} (North-Holland, Amsterdam, 1962).

\bibitem{species1}
T.~Jacobson,
[arXiv:gr-qc/9404039 [gr-qc]].

\bibitem{species2}
J.~D.~Bekenstein,
[arXiv:gr-qc/9409015 [gr-qc]].




\bibitem{Barcelo}
C.~Barcelo, S.~Liberati, S.~Sonego and M.~Visser,
Phys. Rev. D \textbf{83}, 041501 (2011). 

\bibitem{Yang}
H.~Yang, B.~Bonga and Z.~Pan,
Phys. Rev. Lett. \textbf{130}, no.1, 011402 (2023).

\bibitem{Terno}
V.~Baccetti, R.~B.~Mann and D.~R.~Terno,
[arXiv:1904.00506 [gr-qc]].

\bibitem{Edward}
V.~Husain, J.~G.~Kelly, R.~Santacruz and E.~Wilson-Ewing,
Phys. Rev. Lett. \textbf{128}, no.12, 121301 (2022).

\bibitem{QG_con}
D.~Oriti, D.~Pranzetti and L.~Sindoni,
Phys. Rev. Lett. \textbf{116}, no.21, 211301 (2016). 

\bibitem{Oriti}
D.~Oriti,
Class. Quant. Grav. \textbf{33}, no.8, 085005 (2016).


\bibitem{Compere}
G.~Comp\`ere, 
\textit{Advanced Lectures on General Relativity} (Springer, 2019)

\bibitem{Wald}
R.~M.~Wald,
Phys. Rev. D \textbf{48} (1993) no.8, R3427-R3431.


\bibitem{SY}
S.~i.~Sasa and Y.~Yokokura,
Phys. Rev. Lett. \textbf{116}, no.14, 140601 (2016).

\bibitem{SSY}
S.~i.~Sasa, S.~Sugiura and Y.~Yokokura,
Phys. Rev. E \textbf{99}, no.2, 022109 (2019).

\bibitem{HKY}
P.~M.~Ho, H.~Kawai and Y.~Yokokura,
JHEP \textbf{01}, 019 (2022).

\bibitem{Page}
D.~N.~Page,
Phys. Rev. Lett. \textbf{71}, 1291-1294 (1993).

\end{thebibliography}
\end{document}